\documentclass[aip, jmp, amsmath, amssymb, preprint]{revtex4-2}
\usepackage{natbib}
\usepackage[pdftex]{graphicx}% Include figure files
\usepackage{dcolumn}% Align table columns on decimal point
\usepackage{bm}% bold math
\usepackage{color}
\usepackage{url}

\newcommand{\TE}{{\cal T}}
\begin{document}

\preprint{AIP/123-QED}

\title[Assessing transfer entropy from biochemical data]{Assessing transfer entropy from biochemical data}

\author{Takuya Imaizumi, Nobuhisa Umeki$^{1)}$, Ryo Yoshizawa$^{1)}$, Tomoyuki Obuchi$^{2)}$, Yasushi Sako$^{1)}$, and Yoshiyuki Kabashima$^{3)}$}

\affiliation{Department of Mathematical and Computing Science, Tokyo Institute of Technology, 2-12-1 Ookayama, Meguro-ku, Tokyo 152-8550 Japan\\
$^{1)}$Cellular Informatics Laboratory, RIKEN Cluster for Pioneering Research, 2-1 Hirosawa, Wako 351-0198, Saitama, Japan\\
$^{2)}$Department of Systems Science, Kyoto University, 36-1 Yoshida-Honmachi, Sakyo-ku, Kyoto 606-8501, Japan\\
$^{3)}$The Institute for Physics of Intelligence, The University of Tokyo, 7-3-1 Hongo, Bunkyo-ku, Tokyo 113-0033, Japan}
\date{\today}

\begin{abstract}
We address the problem of evaluating the transfer entropy (TE) produced by biochemical reactions from experimentally measured data.
Although these reactions are generally non-linear and non-stationary processes making it challenging to achieve accurate modeling,
Gaussian approximation can facilitate the TE assessment only by estimating covariance matrices using multiple data obtained from 
simultaneously measured time series representing the activation levels of biomolecules such as proteins. 
Nevertheless, the non-stationary nature of biochemical signals makes it difficult to theoretically 
assess the sampling distributions of TE, which are necessary for evaluating the statistical confidence and 
significance of the data-driven estimates. 
We resolve this difficulty by computationally assessing the sampling distributions using techniques from computational statistics. 
The computational methods are tested by using them in analyzing data generated from a theoretically 
tractable time-varying signal model, which leads to the development of a method to screen only statistically significant estimates. 
The usefulness of the developed method is examined by applying it to 
real biological data experimentally measured from the ERBB-RAS-MAPK system 
that superintends diverse cell fate decisions. A comparison between cells containing wild-type and mutant proteins 
exhibits a distinct difference in the time evolution of TE
while apparent difference is hardly found in average profiles of the raw signals. 
Such comparison may help in unveiling important pathways of biochemical reactions. 
\end{abstract}
\maketitle

\section{Introduction}
Cells and their processes are regulated by interactions with biomolecules via chemical reactions. 
Considerable effort has been made to properly understand the biomolecules and their interactions, 
which consititutes a tremendous amount of knowledge \cite{Cell}.
Particularly in the last two decades, chains or  
cascades of reactions have gained increasing attention owing to their potential role in understanding living things as {\em systems} \cite{Iman2018}. 
Various reactions in cells are mathematically modeled using nonlinear differential equations and 
Monte Carlo simulations among others. In addition, large reaction pathways composed of tens or hundreds of
components are drawn as graphs. 

These research efforts have made great progress in understanding the mechanism 
of cell function control. However, there are still shortcomings. 
Although fundamental reactions are modeled by differential equations precisely, 
assessing their significance in the reaction cascade is non-trivial.
An increase in the activation levels of enzymes indicates that certain functions are emerging.
Nevertheless, one cannot quantify
the significance of the reactions by only examining solutions of the differential equations as relevant activation levels can differ from enzyme to enzyme. 
In addition, biochemical reactions are dynamical processes, which indicates that the temporal alteration of signals is 
an important medium for transferring relevant information. Therefore, after the reaction pathways are plotted as graphs, 
there is a need to clarify the timing and how large the transmitted information %being transmitted 
is to properly control cell function.  

For compensating these deficiencies, we focus on the {\em transfer entropy (TE)} \cite{Schreiber2000,Barnett2009}. TE is a universal measure of the 
causal relationship between two time series. It is defined by a functional of the joint distributions of two time series. 
This makes it possible to quantify causal significance in a unified manner, regardless of the physical mechanism
that generates the time series, by representing the generation process in probabilistic models.  
Some empirical studies support its effectiveness in detecting and characterizing causal relations in complex systems~\cite{10.5555/3036153,doi:10.1142/S0218127407017628}.
TE has been used for analyzing information processing in various biological 
organisms from nervous systems \cite{Wibral2010,Olav2012,Ursino2020,Honey10240,PhysRevE.83.051112,PhysRevE.86.066211,SUN201449,10.1162/netn_a_00092,10.1371/journal.pcbi.1008054} to gene regulatory networks \cite{tung2007inferring,10.1007/978-3-319-03844-5_44,castro2019gene,Kim2020} owing to its universality and effectiveness. 

Generally, there are two ways to assess TE: {\em model-based} and {\em data-driven} approaches. 
In the model-based approach, the generation processes of the time series are modeled using high-dimensional joint distributions or their equivalent expressions, in which the TE is evaluated analytically or numerically. 
Although this approach can potentially enable the exact assessment of TE for given models, 
accurately modeling the generation process of actual systems is difficult, which practically limits its application 
range to the analysis of theoretical models \cite{GarciaMichel2021}. 
Meanwhile, in the data-driven approach, the TE is assessed directly from experimentally measured data, which avoids the difficulty in accurately modeling actual systems. 
However, the data-driven approach requires a sufficient number of simultaneously measured data
to substitute the joint distribution of the objective time series as samples. 
Hence, technical constraints have restricted its application mostly to nervous systems, in which large-scale simultaneous 
measurement is possible and the required sample sizes are relatively small because of the stationary nature of the nervous systems 
\cite{Wibral2010,Olav2012,Ursino2020,Honey10240,PhysRevE.83.051112,PhysRevE.86.066211,SUN201449,10.1162/netn_a_00092,10.1371/journal.pcbi.1008054}. 
However, the recent advancement and development of measurement techniques has greatly raised the ability of 
simultaneously measuring the activation levels of biomolecules with high space and time resolutions \cite{Nakamura2017,Yoshizawa2021}. 
This may make it possible to assess TE of non-stationary biochemical interactions
in cells from measured data.  

In such view on the current situation, we explore the possibilities and limitations 
for assessing TE from measured data of biochemical reactions. 
Generally, the assessment of TE is a computationally demanding task involving high-dimensional integrals. 
For overcoming this difficulty, we employ an approximation by Gaussian models, in which 
the assessment can be performed efficiently by estimating 
the covariance matrix of the time series. 
However, it is difficult to theoretically 
assess the sampling distributions of TE, which are indispensable for evaluating statistical confidence and significance 
of the data-driven estimates, due to the non-stationary nature of biochemical signals.
We overcome this difficulty by 
developing a method to screen only statistically significant estimates 
based on techniques from computational statistics, the utility of which 
is tested by the application to a theoretically tractable time-varying signal model. 
In addition, we employ the method for analyzing data experimentally measured from 
the ERBB-RAS-MAPK system of actual cells. Distinct
difference in the time alteration of TE is found in the comparison between cells containing 
wild-type and mutant proteins, whereas there is no apparent difference in the average 
profiles of raw time series. Such comparison may be useful in identifying unknown pathways of biochemical reactions. 

This paper is organized as follows: we review the definition of TE and how it can be computed under 
the Gaussian approximation in Section 2. In addition, we present 
methods for assessing the statistical confidence and significance of 
the inferred results using techniques of computational statistics. 
In Section 3, the utility of the methods is examined by applying them to data from a theoretically tractable time-varying signal model.
Based on the results obtained for the theoretically tractable model, 
we propose a method for screening only statistically significant estimates of TE. 
In Section 4, real-world data from the ERBB-RAS-MAPK system are examined using the method. 
The final section presents the summary and discussion of the study. 

\section{Assessment of the transfer entropy from measured data}
\subsection{Transfer entropy}
Here, we briefly review the definition of TE \cite{Schreiber2000,Barnett2009}.
We use a conventional matrix-vector notation in which the bold type denotes column vectors and 
the uppercase type represents matrices or random variables. 
The symbol $\top$ indicates the matrix-vector transpose, such that the (column) vector $\bm{x} \in \mathbb{R}^n$ is represented as $\bm{x}= (x_1,\ldots, x_n)^\top$.
Given multiple vectors such as $\bm{x} \in \mathbb{R}^n$, $\bm{y} \in \mathbb{R}^{m}$, $\bm{z} \in \mathbb{R}^{\ell}$ and $\ldots$, 
their concatenation is expressed as $\bm{x} \oplus \bm{y} \oplus \bm{z} \oplus \ldots 
= (x_1, \ldots, x_n, y_1, \ldots, y_m, z_1, \ldots, z_\ell, \ldots)^\top$.  
The notation ${\mathcal N}(\bm{\mu}, \Sigma)$ denotes a Gaussian distribution, where $\bm{\mu}$ is the mean vector and $\Sigma$ is the covariance matrix. 

We suppose two interdependent discrete time stochastic processes $X_t$ and $Y_t$ $(t=0,1,\ldots, T)$, 
and denote $\bm{X} \equiv (X_T, X_{T-1}, \ldots, X_0)^\top$ and 
$\bm{Y} \equiv (Y_T, Y_{T-1}, \ldots, Y_0)^\top$.
We also use the notation $\bm{X}_t^{(p)} \equiv (X_t, X_{t-1}, \ldots, X_{t-p+1})^\top$ and 
$\bm{Y}_t^{(q)} \equiv (Y_t, Y_{t-1}, \ldots, Y_{t-q+1})^\top$. 
Let us assume that $\bm{X}$ and $\bm{Y}$ follow 
a joint distribution $Q(\bm{x}, \bm{y})$. 
Under this setup, the uncertainty of $X_t$ is assessed by {\em (differential) entropy} 
\begin{eqnarray}
H(X_t) = - \int dx_t Q(x_t) \ln Q(x_t), 
\label{entropy}
\end{eqnarray}
and that given $\bm{X}_{t-1}^{(p)}$ is quantified by {\em conditional entropy}
\begin{eqnarray}
H(X_t|\bm{X}_{t-1}^{(p)}) = -\int  d\bm{x}_{t-1}^{(p)} dx_t Q(\bm{x}_{t-1}^{(p)}) Q(x_t|\bm{x}_{t-1}^{(p)}) \ln Q(x_t|\bm{x}_{t-1}^{(p)}), 
\label{cond_entropy}
\end{eqnarray}
where $Q(x_t)$, $Q(\bm{x}_{t-1}^{(p)})$, and $Q(x_t|\bm{x}_{t-1}^{(p)})$ represent the marginal and conditional distributions 
with respect to relevant variables. All of these distributions are reduced from the joint distribution $Q(\bm{x}, \bm{y})$, 
and similar notations are employed hereafter without explanations. 
$H(X_t|\bm{X}_{t-1}^{(p)}) $ means the remaining uncertainty of $X_t$ on average given $\bm{X}_{t-1}^{(p)}$. 
Therefore, the amount of the reduction of the uncertainty 
\begin{eqnarray}
I(X_t;\bm{X}_{t-1}^{(p)}) = H(X_t) - H(X_t|\bm{X}_{t-1}^{(p)}), 
\end{eqnarray}
which is referred to as {\em mutual information} between $X_t$ and $\bm{X}_{t-1}^{(p)}$
and can be shown to be non-negative, is interpreted as the amount of information conveyed from 
the past $p$ states $\bm{X}_{t-1}^{(p)} = (X_{t-1}, ..., X_{t-p})^\top $ to the current state $X_t$. 

Extending the notion of mutual information, {\em transfer entropy (TE)} from $Y$ to $X$ 
at time $t$ with lags $q$ and $p$ is defined as
\begin{eqnarray}
\TE_{Y\to X}^{(q), (p)} (t) &=& I(X_t;\bm{X}_{t-1}^{(p)}, \bm{Y}_{t-1}^{(q)}) - I(X_t;\bm{X}_{t-1}^{(p)}) \cr
&=& H(X_t|\bm{X}_{t-1}^{(p)}) - H(X_t|\bm{X}_{t-1}^{(p)}, \bm{Y}_{t-1}^{(q)}), 
\label{TEdefinition1}
\end{eqnarray}
and from $X$ to $Y$ as
\begin{eqnarray}
\TE_{X\to Y}^{(p), (q)} (t) &=& I(Y_t;\bm{X}_{t-1}^{(p)}, \bm{Y}_{t-1}^{(q)}) - I(Y_t;\bm{Y}_{t-1}^{(q)}) \cr
&=& H(Y_t|\bm{Y}_{t-1}^{(q)}) - H(Y_t|\bm{X}_{t-1}^{(p)}, \bm{Y}_{t-1}^{(q)}).
\label{TEdefinition2}
\end{eqnarray}
Equation (\ref{TEdefinition1}) indicates how much information is increased regarding $X_t$ or how much uncertainty about $X_t$ is reduced 
by providing $\bm{Y}_{t-1}^{(q)}$ on top of $\bm{X}_{t-1}^{(p)}$, and similarly for (\ref{TEdefinition2}). These mean that TE stands for the significance of past states of one variable in 
predicting the current state of the other.
In addition, the TE is asymmetric between $X$ and $Y$ unlike mutual information. 
Therefore, TE is employed as a useful measure of information flow that quantifies the causal relationship between two time series.

\subsection{TE for Gaussian models}
TE assessment is computationally demanding even though it is expressed in compact forms, such as (\ref{TEdefinition1}) and (\ref{TEdefinition2}). 
This is because the computational complexity grows exponentially with respect to $p+q+1$ when numerically evaluating (\ref{TEdefinition1}) and (\ref{TEdefinition2}), which can be infeasible 
even for the minimum lags of $p=q=1$ in practical situations. 
However, when $Q(\bm{x}, \bm{y})$ is a multivariate Gaussian distribution,  
the assessment becomes feasible as entropy and conditional entropy can be analytically 
evaluated using covariance matrices for the Gaussian model. 

For explaining this more precisely, we introduce the notation $\Sigma(\bm{U}) $ to generally 
express the covariance matrix of the random vector $\bm{U}$. 
%On the other hand, 
In addition, we use the notation $\Sigma(\bm{U}, \bm{V})$ to denote 
the cross-covariance matrix between $\bm{U}$ and $\bm{V}$, 
which is composed of their cross-covariances ${\rm cov}(U_i, V_\alpha)$. 
%In addition, we define 
These provide the {\em partial covariance} \cite{Barnett2009} of $\bm{U}$ given $\bm{V} \oplus \bm{W} \oplus \ldots$ as
\begin{eqnarray}
&&\Sigma(\bm{U}| \bm{V} \oplus \bm{W} \oplus \ldots)
= \Sigma(\bm{U}) \cr
&&\phantom{===}-\Sigma(\bm{U}, \bm{V}\oplus \bm{W} \oplus \ldots) \Sigma(\bm{V} \oplus \bm{W} \oplus \ldots)^{-1}.  
\Sigma(\bm{U}, \bm{V}\oplus \bm{W} \oplus  \ldots)^\top.
\label{pratialCovariance}
\end{eqnarray}
When $Q(\bm{x}, \bm{y})$ is given as a multivariate Gaussian distribution, the properties of the Gaussian random variables 
yield a formula 
\begin{eqnarray}
H(X_t|\bm{X}_{t-1}^{(p)})  = \frac{1}{2}\ln \left (\Sigma(X_t|\bm{X}_{t-1}^{(p)}) \right ) + \frac{1}{2}\ln (2\pi {\color{black}e}), 
\label{GaussianEntropy}
\end{eqnarray}
where $\Sigma(X_t|\bm{X}_{t-1}^{(p)}) = \Sigma(X_t) - \Sigma(X_t, \bm{X}_{t-1}^{(p)}) \Sigma(\bm{X}_{t-1}^{(p)})^{-1} 
\Sigma(X_t, \bm{X}_{t-1}^{(p)})^\top$ is the variance of $X_t$ conditioned by $\bm{X}_{t-1}^{(p)}$. 
{\color{black} 
The derivation of this formula is presented in Appendix \ref{GaussianEntropyDeriv}. 
}
Similarly, we also obtain  
\begin{eqnarray}
H(X_t|\bm{X}_{t-1}^{(p)}, \bm{Y}_{t-1}^{(q)})  = \frac{1}{2}\ln \left (\Sigma(X_t|\bm{X}_{t-1}^{(p)} \oplus \bm{Y}_{t-1}^{(q)} )\right ) + \frac{1}{2}\ln (2\pi {\color{black}e}). 
\end{eqnarray}
These formulas provide an expression of TE for the Gaussian time series as
\begin{eqnarray}
\TE_{Y\to X}^{(q), (p)} (t)  = 
\frac{1}{2}\ln \left (\frac{\Sigma(X_t|\bm{X}_{t-1}^{(p)}) }{\Sigma(X_t|\bm{X}_{t-1}^{(p)} \oplus \bm{Y}_{t-1}^{(q)} )} 
\right ), 
\label{GaussianTE}
\end{eqnarray}
and similarly for $\TE_{X\to Y}^{(p), (q)} (t)$. 

Several issues are to be noted with this formula. 
The first is about the implications of (\ref{GaussianTE}). 
This formula indicates that TE is determined using only covariances between the two time series irrespectively of their averages.  
This implies that the primary media of information transfer are not the average profiles of the observed signals
but their statistical fluctuations, which may be counterintuitive. 
Nevertheless,  in the framework of the Gaussian approximation, 
the covariances are determined by the Hessian of $-\ln Q(\bm{x}, \bm{y})$ 
around the averages for general joint distributions $Q(\bm{x}, \bm{y})$. 
This means that the average profile of signals is not a unique but still a major factor for determining TE. 
The second is about the cost for computation. Given the covariance matrix 
$\Sigma(X_t \oplus \bm{X}_{t-1}^{(p)} \oplus \bm{Y}_{t-1}^{(q)})$, 
the computational cost for assessing (\ref{GaussianTE}) increases as $O((p+q)^{3})$ 
since the most computationally intensive part is the assessment of the matrix inversion 
$\Sigma(\bm{X}_{t-1}^{(p)} \oplus \bm{Y}_{t-1}^{(q)})^{-1}$. 
In most cases, this is computationally feasible as long as $p$ and $q$ are $O(1)$.  
The third issue is that, as mentioned in \cite{Barnett2009}, TE is equivalent to the Granger causality \cite{Granger1969}, which has been extensively studied since 1970s, for time series generated by multi-variate 
autoregressive (MVAR) models. However, in the framework of the Granger causality, 
we must introduce many assumptions about how to describe the time series 
by MVAR models, which becomes nontrivial, particularly when handling non-stationary time series.
In contrast, the TE framework is ``model agnostic'' \cite{Barnett2009} as
(\ref{GaussianTE}) can be assessed directly from the covariances, for which we need few assumptions. 
Therefore, the formula of (\ref{GaussianTE}) would be more user-friendly when 
sufficient knowledge about the data generation process is not available. 
The final issue is concerning the validity of resorting to Gaussian models. 
The appropriateness of modeling time series as Gaussians may be criticized for specific physical generation processes. 
However, in most cases, assessing the TE from the exact formula of (\ref{TEdefinition1}) and (\ref{TEdefinition2})
requires significantly heavy computations for non-Gaussian models; therefore, Gaussian models are practically unique choices. 
Further, fortunately, it is known that even in non-Gaussian cases
non-zero values of (\ref{GaussianTE}) imply that non-zero values of the true TE given by (\ref{TEdefinition1}) \cite{Marinazzo2008}. 
For these reasons, we use the formula of (\ref{GaussianTE}) for assessing TE.

\subsection{Assessment of TE from data and its statistical significance}
Following the above argument, TE can be assessed from samples of time 
series ${\mathcal D}_M = \{\bm{x}_\mu, \bm{y}_\mu\}_{\mu=1}^M$ by 
substituting $\Sigma(X_t \oplus \bm{X}_{t-1}^{(p)} \oplus\bm{Y}_{t-1}^{(q)})$ with its estimate from ${\mathcal D}_M$ 
in evaluating (\ref{GaussianTE}). A natural estimator of $\Sigma(X_t \oplus \bm{X}_{t-1}^{(p)} \oplus \bm{Y}_{t-1}^{(q)})$ is 
\begin{eqnarray}
\hat{\Sigma}(X_t \!\oplus \!\bm{X}_{t-1}^{(p)} \!\oplus\!  \bm{Y}_{t-1}^{(q)})
\! = \!\frac{1}{M-1} \sum_{\mu=1}^M (\bm{z}_{t,\mu}^{(p), (q)} \!-\!\overline{\bm{z}_t^{(p),(q)}})
(\bm{z}_{t,\mu}^{(p), (q)} \!-\!\overline{\bm{z}_t^{(p),(q)}})^\top, 
\label{unbiasedSigma}
\end{eqnarray}
where $\bm{z}_{t,\mu}^{(p), (q)}  \equiv (x_{t,\mu}, x_{t-1, \mu}, \ldots, x_{t-p, \mu}, y_{t-1, \mu}, \ldots, y_{t-q, \mu})^\top$
represents the concatenation of $\mu$-th samples $x_{t, \mu}$, $\bm{x}_{t-1, \mu}^{(p)}$, and $\bm{y}_{t-1, \mu}^{(q)}$, 
and $\overline{\bm{z}_t^{(p),(q)}} \equiv M^{-1} \sum_{\mu=1}^M\bm{z}_{t,\mu}^{(p), (q)} $, respectively. 
This is a consistent and unbiased estimator of $\Sigma(X_t \oplus \bm{X}_{t-1}^{(p)} \oplus \bm{Y}_{t-1}^{(q)})$, which guarantees that 
(\ref{unbiasedSigma}) converges to the true
covariance %$\Sigma(X_t \oplus \bm{X}_{t-1}^{(p)} \oplus\bm{Y}_{t-1}^{(q)})$ 
as $M\to \infty$ and the average of (\ref{unbiasedSigma}) with respect to the generation of ${\mathcal D}_M$ 
accords with %$\Sigma(X_t \oplus \bm{X}_{t-1}^{(p)} \oplus \bm{Y}_{t-1}^{(q)})$ 
it for finite $M$. 
The assessment of (\ref{GaussianTE}) using (\ref{unbiasedSigma}) 
also offers a consistent estimator of $\TE_{Y\to X}^{(q), (p)} (t)$ under Gaussian assumptions.
However, this is generally biased. This is natural because $\TE_{Y\to X}^{(q), (p)} (t)$ is a non-negative quantity by nature. 
Therefore, even if $X_t$ and $\bm{Y}_{t-1}^{(q)}$ are statistically independent, yielding $\TE_{Y\to X}^{(q), (p)} (t)=0$, the statistical fluctuations in (\ref{unbiasedSigma}) 
always results in positive values for the estimates of $\TE_{Y\to X}^{(q), (p)} (t)$. 
In addition, the convergence rate of the covariance matrix estimator of (\ref{unbiasedSigma}) 
is rather slow \cite{Ledoit2004}. This makes it challenging to analytically evaluate the sampling distributions of the estimates of TE, 
which are indispensable for assessing the statistical confidence and significance of the inferred results,  although 
it is known that the maximum likelihood estimator of TE will asymptotically 
have a $\chi^2$-distribution under the null hypothesis in which the true TE vanishes, in the case of stationary time series \cite{Granger1963,Whittle1953}. 

For practically overcoming this difficulty, we employ computational methods known as {\em bootstrapping} 
\cite{Efron1993}. We construct the sampling distributions in the following ways depending on the purpose:  

\paragraph{For confidence interval:}
Suppose that an estimate 
$\hat{\TE}_{Y\to X}^{(q), (p)} (t)$ of (\ref{GaussianTE}) 
is evaluated for a given dataset ${\mathcal D}_M$ by substituting  $\Sigma(X_t \oplus\bm{X}_{t-1}^{(p)} \oplus \bm{Y}_{t-1}^{(q)})$ with 
$\hat{\Sigma}(X_t \oplus \bm{X}_{t-1}^{(p)} \oplus \bm{Y}_{t-1}^{(q)})$ of (\ref{unbiasedSigma}). 
We need to evaluate the degree of statistical fluctuations in $\hat{\TE}_{Y\to X}^{(q), (p)} (t)$, which 
is inevitable owing to the finiteness of the sample size $M$. 
For this purpose, 
we handle $\hat{\Sigma}(X_t \oplus \bm{X}_{t-1}^{(p)} \oplus \bm{Y}_{t-1}^{(q)})$ as if it is the true covariance matrix and 
generate a new sample set of size $M$ by independently drawing samples from an identical distribution 
${\mathcal N} \left (0, \hat{\Sigma}(X_t \oplus \bm{X}_{t-1}^{(p)} \oplus\bm{Y}_{t-1}^{(q)}) \right )$. 
This provides a surrogate estimate  of TE $\hat{\TE}_{Y\to X}^{*(q), (p)} (t)$  using  
$\hat{\Sigma}^*(X_t \oplus \bm{X}_{t-1}^{(p)} \oplus \bm{Y}_{t-1}^{(q)})$, which is the covariance matrix estimated from the new sample set. 
We repeat these procedures for $B(\gg 1)$ 
times, which results in an empirical distribution of $\hat{\TE}_{Y\to X}^{*(q), (p)} (t)$. 
We construct the $100(1-\alpha)$\% $(0<\alpha <1)$ confidence interval (CI) by specifying $50\alpha$ and 
$100 -50 \alpha$ percentile points of the empirical distribution to evaluate the degree of statistical fluctuation of the estimate. 
The same procedure is performed for $\hat{\TE}_{X\to Y}^{*(p), (q)} (t)$ as well. 

\paragraph{For significance threshold:}
When estimating $\hat{\TE}_{Y\to X}^{(q), (p)} (t)$, it always takes a non-negative value even when 
the true TE ${\TE}_{Y\to X}^{(q), (p)} (t)$ vanishes. 
We need to evaluate the distribution of $\hat{\TE}_{Y\to X}^{(q), (p)} (t)$ for the null hypothesis $H_0: {\TE}_{Y\to X}^{(q), (p)} (t) =0$ 
to distinguish the obtained value from those of chance levels. 
To construct the distribution, we define the covariance matrix $\Sigma_{H_0} (X_t \oplus \bm{X}_{t-1}^{(p)} \oplus \bm{Y}_{t-1}^{(q)})$ 
for $H_0$ by making all elements of the cross-covariance between $X_t \oplus \bm{X}_{t-1}^{(p)}$ and $\bm{Y}_{t-1}^{(q)}$
vanish in $\hat{\Sigma}(X_t \oplus \bm{X}_{t-1}^{(p)} \oplus \bm{Y}_{t-1}^{(q)})$, 
keeping the other elements fixed. 
Based on this, we generate a new sample set of size $M$ by independently drawing samples from an identical distribution 
${\mathcal N} \left (0, \Sigma_{H_0} (X_t \oplus \bm{X}_{t-1}^{(p)} \oplus \bm{Y}_{t-1}^{(q)}) \right )$. 
This provides an estimate of TE $\hat{\TE}_{Y\to X}^{\# (q), (p)} (t)$  using  
$\hat{\Sigma}^\# (X_t \oplus \bm{X}_{t-1}^{(p)} \oplus \bm{Y}_{t-1}^{(q)})$, 
which is the covariance matrix estimated from the new sample set. 
We repeat these procedures for $B(\gg 1)$ 
times, which provides an empirical distribution of $\hat{\TE}_{Y\to X}^{\# (q), (p)} (t)$. 
The $100(1-\alpha)$ $(0< \alpha < 1)$ percentile point of the empirical distribution is used as 
the significance threshold (ST) to distinguish the estimated value from those of chance levels with 
the significance level of $100\alpha \%$.
The same procedure is performed for $\hat{\TE}_{X\to Y}^{\# (p), (q)} (t)$ as well.

%\vskip\baselineskip

These methods are very simple in terms of technical aspects. 
The necessary procedure is just repeating the  evaluation of the TE from resampled data many times.  
The computational burden for repetition can prevent the execution of these methods in cases where 
a considerable amount of computation is required in obtaining a single estimate. 
However, in the current case, the computational cost for assessing 
TE is only $O((p+q)^3)$ under the Gaussian approximation, 
which would not be an obstacle for the execution. 
Nevertheless, the validity of using not the true distribution but 
the empirical distribution 
provided by the estimated covariance matrices for assessing the sampling distributions, following the {\em plug-in principle}
\cite{Efron1993}, is debatable.  
In the next section, we examine this issue by applying the methods to a time-varying signal model, in which 
the theoretical evaluation of TE is tractable.

\section{Testing the developed method using the theoretically tractable model}
As a simple but non-trivial example for which the theoretical assessment of TE is possible, 
we consider a $d$-dimensional time-varying state space model \cite{Durbin2001}, which is expressed as  
\begin{eqnarray}
&&\bm{u}_{t} = F_t \bm{u}_{t-1} + \bm{\mu}_{t} + \bm{\xi}_{ t}^{\rm U}, \label{TVSSM1} \\
&& \bm{v}_{t} = G \bm{u}_{t} + \bm{\xi}_{t}^{\rm V} \label{TVSSM2}
\end{eqnarray}
where $t = 1, \ldots, T$, and $\bm{u}_t \in \mathbb{R}^{d (\ge 2)}$ 
and $\bm{v}_{t} \in \mathbb{R}^{2}$, respectively. $\bm{\xi}_{t}^{\rm U} \in \mathbb{R}^d$ and $\bm{\xi}_{ t}^{\rm V} \in \mathbb{R}^2$ 
are the Gaussian noise vectors, both of which are independent in time. 
Equation (\ref{TVSSM1}) describes how the state vector $\bm{U}_t$ evolves in time, being subject to 
time-varying parameters $F_t \in \mathbb{R}^{d\times d}$ and $\bm{\mu}_t\in \mathbb{R}^{d}$.
Meanwhile, (\ref{TVSSM2}) defines the measurement process of $\bm{U}_t$. 
We consider that the first and last components of $\bm{U}_t$ are measured, such that 
the components of the resulting vector $\bm{V}_t$ are regarded as $X_t$ and $Y_t$ in Section 2
under the setup of the measurement matrix 
$G=\left (
\begin{array}{ccccc}
1 & 0 &\ldots &  0 & 0 \cr
0  & 0  & \ldots & 0& 1
\end{array}
\right )$. 

Two features are noted for (\ref{TVSSM1}) and (\ref{TVSSM2}):
\begin{itemize}
\item Given a set of the time-varying parameters $F_t$ and $\bm{\mu}_t$, 
the resulting time series $\bm{u}_T, \ldots, \bm{u}_1$ and $\bm{v}_T, \ldots, \bm{v}_1$ are provided 
as linear combinations of $\bm{\xi}_{T}^{\rm U}, \ldots,  \bm{\xi}_{ 1}^{\rm U}$ and 
$\bm{\xi}_{T}^{\rm V}, \ldots, \bm{\xi}_{ 1}^{\rm V}$ added to deterministic time series for a given initial state $\bm{u}_0$. 
This guarantees that the set of time series $\bm{U} = \{\bm{U}_T, \ldots, \bm{U}_1\}$ and $\bm{V} = \{\bm{V}_T, \ldots, \bm{V}_1\}$
follows a joint Gaussian distribution that varies over time. 
% because these are samples have Gaussian distributions, 
\item The mean parameters
$
\bm{\mu}_T,  \ldots, \bm{\mu}_{1}
$ 
do not contribute to statistical fluctuations of $\bm{U}$. 
This indicates that covariances of any components of 
$\bm{U}$ and $\bm{V}$ are independent of the mean parameters. 
\end{itemize}

These features enable us to evaluate $\Sigma(X_t \oplus\bm{X}_{t-1}^{(p)} \oplus \bm{Y}_{t-1}^{(q)})$
and $\Sigma(Y_t \oplus \bm{X}_{t-1}^{(p)} \oplus \bm{Y}_{t-1}^{(q)})$ efficiently using recursive equations, 
which results in the theoretical assessment of TE.  
More specifically, the statistical independence of noise vector $\bm{\xi}_{t}^{\rm U}$ regarding time provides
a recursive equation for computing covariance matrix $C_t^{\rm U} = \mathbb{E}[\bm{U}_t \bm{U}_t^\top ]
- \mathbb{E}[\bm{U}_t]  \mathbb{E}[\bm{U}_t]^\top$, which is expressed as 
\begin{eqnarray}
C_{t+1}^{\rm U} = F_t C_{t}^{\rm U} F_t^\top + \Delta_t^{\rm U}, 
\label{recursive_CU}
\end{eqnarray}
where $\mathbb{E}[\cdots]$ is the average with respect to noise vectors
and $\Delta_t^{\rm U}$ is the covariance matrix of $\bm{\xi}_{t}^{\rm U}$. 
In addition, the covariance matrix of $\bm{U}_t$ between two different times are given as
\begin{eqnarray}
D_{t+m, t}^{\rm U} = \left (\prod_{\tau = t+1}^{t+m} F_\tau \right ) C_t^{\rm U}, 
\label{recursive_DU}
\end{eqnarray}
where
$m=1,\ldots, T-t$ and 
 $D_{t+m, t}^{\rm U} = \mathbb{E}[\bm{U}_{t+m} \bm{U}_t^\top ]
- \mathbb{E}[\bm{U}_{t+m}]  \mathbb{E}[\bm{U}_t]^\top$. 
Subsequently, the covariance matrices of $\bm{V}_t$ are computed as 
\begin{eqnarray}
&&C_{t}^{\rm V} = G C_{t}^{\rm U} G^\top + \Delta_t^{\rm V}, \label{recursive_CV} \\
&&D_{t+m, t}^{\rm V} = G D_{t+m, t}^{\rm U} G^\top, \label{recursive_DV}
\end{eqnarray}
where $C_{t}^{\rm V} =\mathbb{E}[\bm{V}_t \bm{V}_t^\top ]
- \mathbb{E}[\bm{V}_t]  \mathbb{E}[\bm{V}_t]^\top$, 
$D_{t+m, t}^{\rm V} = \mathbb{E}[\bm{V}_{t+m} \bm{V}_t^\top ]
- \mathbb{E}[\bm{V}_{t+m}]  \mathbb{E}[\bm{V}_t]^\top$, and 
$\Delta_t^{\rm V}$ is the covariance matrix of $\bm{\xi}_{t}^{\rm V}$. 
The recursive equations (\ref{recursive_CU}) and (\ref{recursive_DU}), in conjunction with (\ref{recursive_CV}) and (\ref{recursive_DV}), 
offer all components that constitute $\Sigma(\bm{X} \oplus \bm{Y})$
with $O(T^2)$ computational cost.  
Performing this is feasible as long as $p$ and $q$ are $O(1)$. 
Picking up all components of $\Sigma(X_t \oplus \bm{X}_{t-1}^{(p)} \oplus \bm{Y}_{t-1}^{(q)})$
from $\Sigma(\bm{X} \oplus\bm{Y})$ and substituting them into (\ref{GaussianTE}) provide
$\TE_{Y \to X}^{(q),(p)}(t)$, and similarly for $\TE_{X \to Y}^{(p),(q)}(t)$.

\begin{figure}[t]
  \centering
  \includegraphics[width = 10cm]{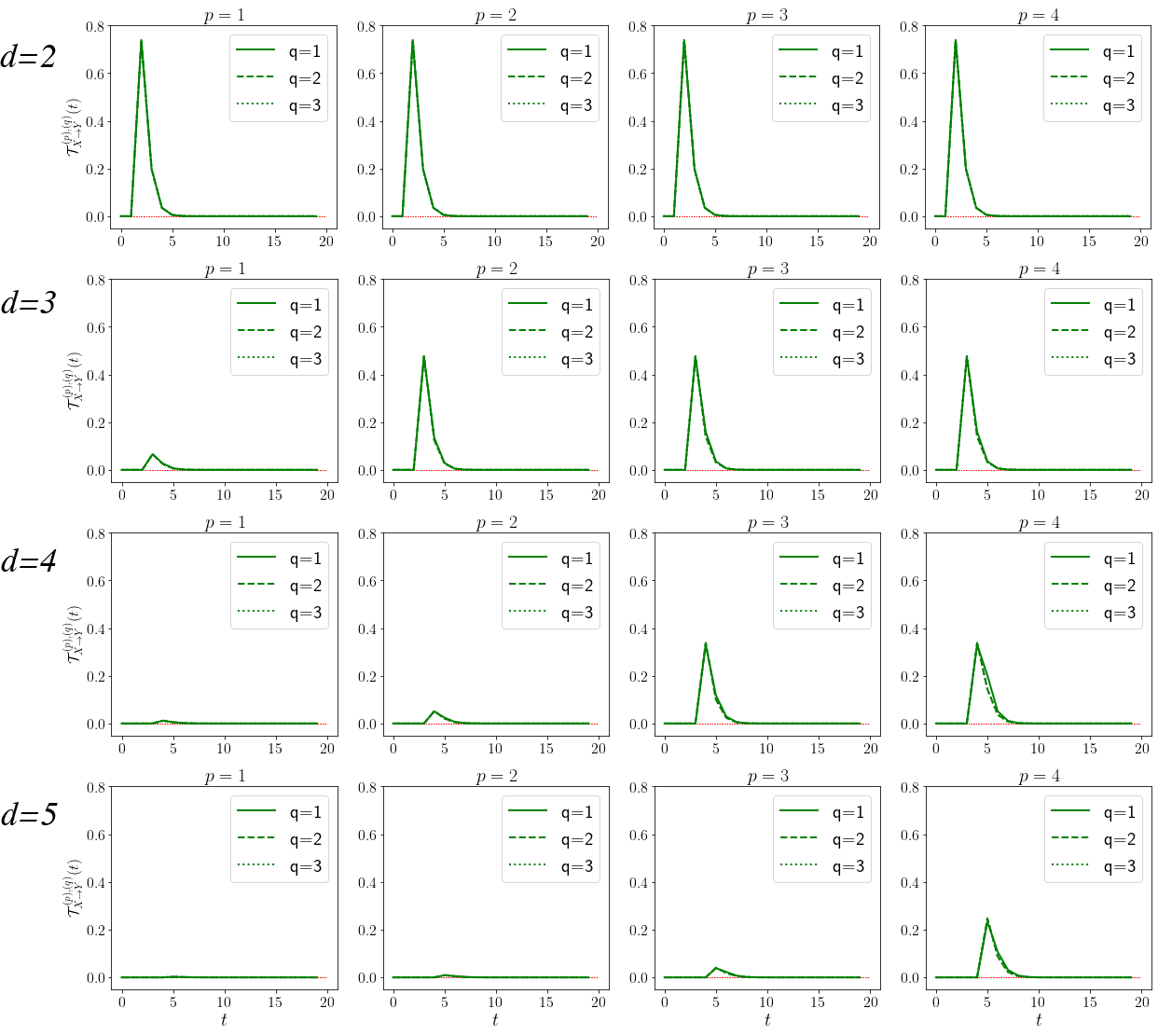}
  \caption{ Theoretically computed $\TE_{{ X} \to { Y}}^{(p), (q)}(t)$ 
  for the system characterized by (\ref{Ft}). 
  % for various combinations of lag parameters $p$ 
  % and $q$ by varying the cascade length $d$ from $2$ to $5$. 
  From top to bottom: The cascade length $d$ varies from $2$ to $5$.  
  From left to right: The lag $p$ of the cause time series $X_t$ changes from $1$ to $4$. 
  See the text for details. 
  \label{fig1}}
\end{figure}

\begin{figure}[t]
  \centering
  \includegraphics[width = 8cm]{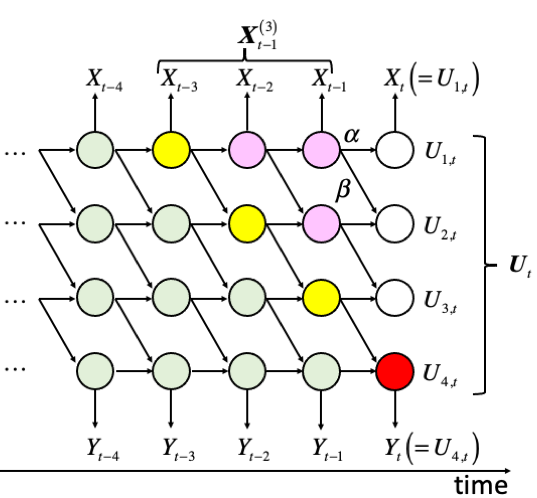}
  \caption{Computational graph of (\ref{Ft})  in the case of $d=4$. 
  \label{fig2}}
\end{figure}

For simulating time series that are simultaneously measured from a cascade of reactions, 
we consider a very simple model, which is given by 
\begin{eqnarray}
F_t = \left (
\begin{array}{ccccc}
\alpha & 0 & \ldots & \ldots & 0 \cr
F_{21}(t) & \alpha & 0 & \ldots & 0\cr
0 & \beta  &  \ddots & \ddots & \vdots \cr
\vdots & \vdots & \ddots & \alpha & 0 \cr
0 & 0 & \ddots & \beta & \alpha
\end{array}
\right ),
\label{Ft}
\end{eqnarray}
where $0<\alpha < 1$ is the decay rate of each component and
$\beta $ is the activation caused by  the reaction with the component of the previous subscript. 
We assume that the first component of $\bm{U}$ triggers the reaction cascade in a short duration $\gamma > 0$, which is 
taken into account by the time-varying matrix element
\begin{eqnarray}
F_{21}(t) = \left \{
\begin{array}{ll}
\delta \exp(-t/\gamma ), & t \ge 0 \cr
0, & t < 0
\end{array}
\right .  
\label{F21} . 
\end{eqnarray}

Indeed, we need to employ nonlinear equations with respect to $\bm{u}_t$ to precisely describe realistic cascades 
of chemical reactions. 
However,  the media for information transfer in the framework of the Gaussian treatment is the statistical 
fluctuations of the state vectors around their averages, 
which justifies the current linearized description at least as a first approximation. 
In addition, the uniform setting of the parameters is not crucial when 
examining how the length of the reaction cascade $d$ 
is reflected in the dependence of TE on the lag parameters $p$ and $q$, which we will focus on in the following discussions.

We set $\alpha = 0.5$ and $\beta = 1$ to ensure that the influence of the trigger of the first component propagates 
forward without decay. Meanwhile, we set $\delta = 5$ and $\gamma =1$, 
$\Delta_{t}^{\rm U} = 0.1^2 \times I_{d\times d}$, and $\Delta_{t}^{\rm V} = 0$ for other parameters, 
where $I_{n\times n}$ generally denotes $n \times n$ identity matrix.  
Figure \ref{fig1} presents the plot of $\TE_{ {X} \to { Y}}^{(p), (q)}(t)$ 
computed from (\ref{recursive_CU})--(\ref{recursive_DV}) for various pairs of lag parameters $p$ and $q$ by varying the length of the reaction cascade $d$ from $2$ to $5$. 
The TE of the reverse direction can also be computed, but it is not presented 
because the statistical independence between the first component of $\bm{U}_t$ and 
the subsequent components in the past state $\bm{U}_{t-m}$ ($m\ge 1$) 
guarantees that $\TE_{{ Y} \to { X}}^{(q), (p)}(t)$ vanishes trivially. 
All the plots indicate that $\TE_{{ X} \to {Y}}^{(p), (q)}(t)$ does not significantly depends on the lag $q$ 
of the {\em effect} time series $Y_t$. 
This is presumably because in the current setup, $\Delta_{t}^{\rm V} = 0$ is set to zero for a relatively small $\alpha$. 
When $\Delta_{ t}^{\rm V}$ is set finite, the dependence of the TE on the lag $q$ of the effect time series varies 
in a non-trivial manner depending on the value of $\alpha$. 

On the other hand, $\TE_{{ X} \to { Y}}^{(p), (q)}(t)$ monotonically increases as the lag $p$ of the {\em cause} time series $X_t$ increases, and almost saturates at $p= d-1$. 
This is because the influence of the first component $U_{1,t}(=X_t)$ of $\bm{U}_t$ reaches the last component 
$U_{d,t}(=Y_t)$ late by the lag of $d-1$
due to to the one-dimensional nature of the reaction cascade, as shown in the computational graph in Figure \ref{fig2}. 
This graph indicates that the prediction accuracy of $Y_t$ (red node) is maximized when 
$U_{d-1, t-1}$ is given in addition to $Y_{t-1}$. 
Unfortunately, $U_{d-1, t-1}$ is a hidden variable and cannot be directly observed. 
However, it is correlated with past states $X_{t-1}(=U_{1, t-1}), X_{t-2}(=U_{1, t-2}), \ldots$
as they share the same ancestors, and the strength of the correlation is maximized by $X_{t-d+1}$ 
as it is connected to $U_{d-1, t-1}$ without decaying through the path of the sequence of yellow nodes in the graph.
Therefore,  $\TE_{{ X} \to { Y}}^{(p), (q)}(t)$  increases as $p$ increases from $1$ to $d-1$. 
However, for $p > d-1$, the information from $X_{t-d}, \ldots, X_{t-p}$ has already been considered 
in $Y_{t-1}(=U_{d,t-1}), \ldots, Y_{t-p+d}(=U_{d, t-p+d})$, which are denoted as light green nodes in the graph, 
and yields little innovative gain on top of $\bm{Y}_{t-1}^{(q)}$ 
in predicting $Y_t$. Therefore, $\TE_{{ X} \to { Y}}^{(p), (q)}(t)$ hardly increases for $p> d-1$, 
which is the reason for the saturation at $p=d-1$.
{\color{black} Such dependence of TE on the lag parameter of the cause time series
does not change qualitatively unless decay rates $\alpha$ and $\beta$ vary  depending on sites
significantly around the above-mentioned values. 
This, in conjunction with 
rough knowledge about the time scale of elemental reactions, may be useful  
for characterizing the length of the reaction cascade. }

%{\color{black} Although the above interpretation is justified only under the assumption 
%that all reactions have almost the same damping rate $\alpha$, 
%such dependence of TE on the lag parameter of the cause time series, in conjunction with 
%rough knowledge about the time scale of elemental reactions, may be useful  
%for characterizing the length of the reaction cascade. }

\begin{figure}[t]
  \centering
  \includegraphics[width = 12cm]{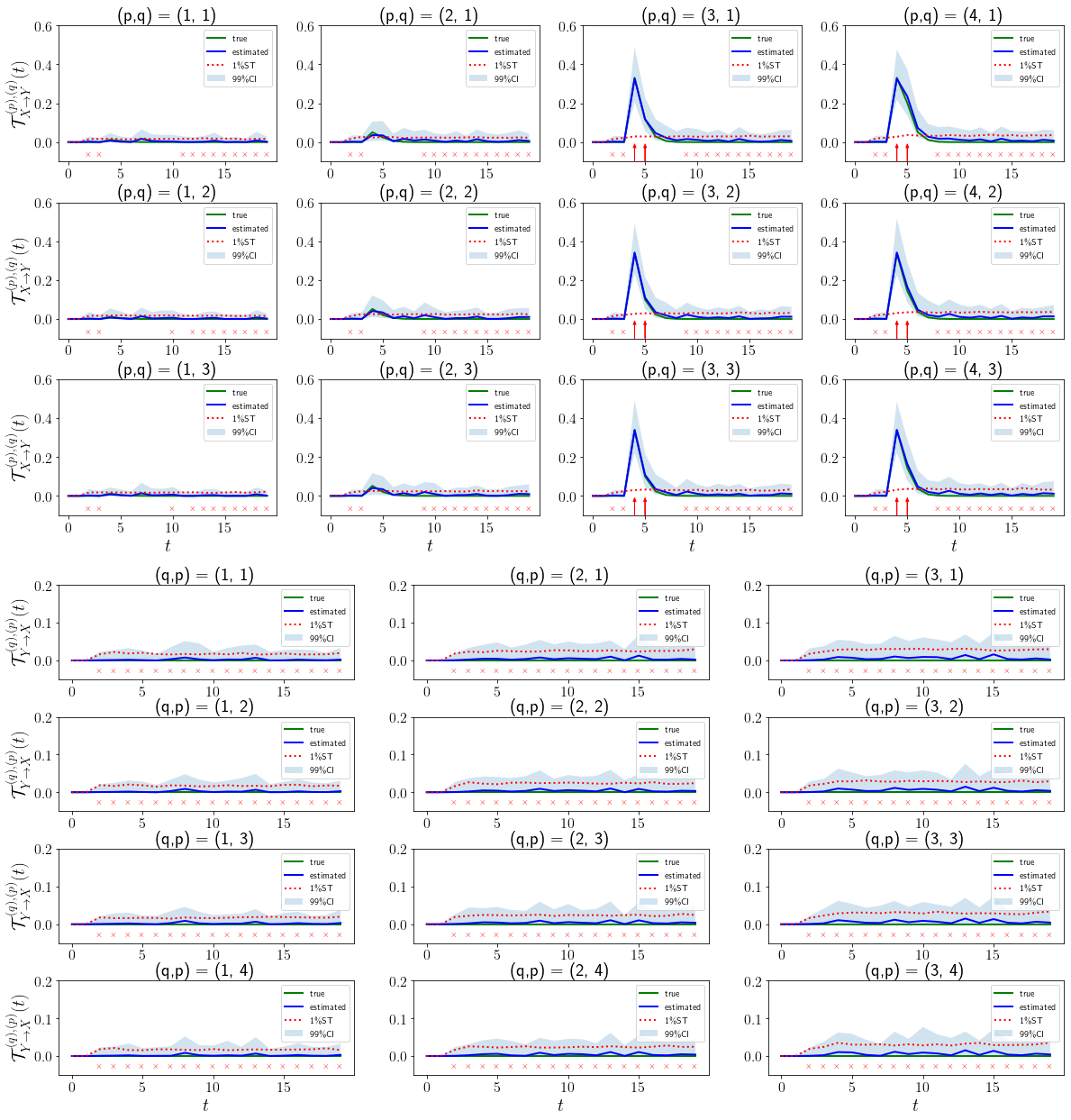}
  \caption{ (a): $\TE_{{ X} \to {Y}}^{(p), (q)}(t)$ assessed from 200 samples for the $d=4$ case presented in Figure \ref{fig1}. 
  (b): That for $\TE_{{ Y} \to {X}}^{(q), (p)}(t)$. 
  In both panels, blue and green full lines stand for the estimated and true values, respectively. 
  The true values of $\TE_{{ Y} \to {X}}^{(q), (p)}(t)$ are constantly zero in this setting. 
  Light blue areas and red dotted lines represent 99\%CIs and 1\%STs. 
  % The true values of $\TE_{{ Y} \to {X}}^{(q), (p)}(t)$, which are constantly zero, are also plotted for reference.  
  %Although the upper limit of the 99\%CI exceeds the true value of the TE (green) at every point, 
  The lower limit of the 99\%CI exceeds the true value of the TE 
  at points marked by red crosses, which may 
  overestimate TE. However, one can screen statistically significant 
  estimates (marked by red arrows) by accepting only cases in which the lower limit of the CI is larger than the 1\%ST. 
  \label{fig3}}
\end{figure}

{ Figure \ref{fig3} shows $\TE_{{ X} \to {Y}}^{(p), (q)}(t)$  assessed 
from $N_{\rm samp} = 200$ samples generated by (\ref{TVSSM1}) 
and (\ref{TVSSM2}) for the $d=4$ case presented in Figure \ref{fig1} using the methods discussed in Section 2. 
Estimates of $\TE_{{ Y} \to {X}}^{(q), (p)}(t)$, the true values of which are constantly zero, are also plotted for reference. }
For all plots, the upper limit of 99\%CIs surpassed the true value (green) at every point. 
However, its lower limit is also larger than the true value at points where 
the true values are small 
(marked by red crosses). 
In such cases, the CIs do not cover the true values, leading to inaccurate estimations.

{\color{black} 
Cross covariances in the estimated matrices 
$\hat{\Sigma}(X_t \oplus \bm{X}_{t-1}^{(p)} \oplus\bm{Y}_{t-1}^{(q)})$ 
and $\hat{\Sigma}(Y_t \oplus \bm{X}_{t-1}^{(p)} \oplus\bm{Y}_{t-1}^{(q)})$
are always non-zero even if two time series $\bm{X}$ and $\bm{Y}$ are statistically independent. 
%In addition, TE takes positive values unless all the cross covariances completely vanish. 
This means that TE assessed from samples is always biased positively even %if $\bm{X}$ and $\bm{Y}$ are statistically independent,  
for statistically independent two time series, 
% and similar tendency also holds when statistical correlations of $\bm{X}$ and $\bm{Y}$ are weak.  
% This implies that TE computed from bootstrap samples is likely to be overestimated in comparison with the true value, 
which could lead to a risk of giving false positives in judging the finiteness of TE.  
One approach to reduce the risk is to estimate the covariance matrices extremely accurately by 
collecting a huge number of samples. However, this is difficult to carry out in practice. 
Another approach is to take into account the positive biases in judging the statistical significance. 
In Figure \ref{fig3}, the lower limit of the 99\%CI is smaller than the 1\%ST at all points that are marked by the red crosses, 
indicating that the estimates are not statistically significant in the worst case. 
In other words, statistically significant estimates can be screened under given statistical confidence and significance levels
by accepting only cases where the lower limit of the CI is larger than ST (marked by red arrows). 
This is the method for assessing TE that we propose in this study. 
}

% Another factor that could affect the estimation accuracy of the method is the shortage of the sample size $N_{ \rm samp}$. 
{\color{black} 
The degrees of freedom of covariance matrices $\hat{\Sigma}(X_t \oplus\bm{X}_{t-1}^{(p)} \oplus\bm{Y}_{t-1}^{(q)})$ and  
$\hat{\Sigma}(Y_t \oplus \bm{X}_{t-1}^{(p)} \oplus \bm{Y}_{t-1}^{(q)})$ that are necessary for assessing TE with 
lag parameters $p$ and $q$ are $(p+q+2)(p+q+1)/2$. This is as high as $6$ even for the smallest case of $p=q=1$, and 
grows up to 28 for $p=q=3$. 
As $N_{ \rm samp}$ must be sufficiently larger than the degrees of freedom for accurate estimation of the matrices, 
the sample size should be at least hundreds even if the proposed method is employed. }

\section{Application to data from ERBB-RAS-MAPK system}
The proposed method for assessing TE was applied to real biological data obtained from single living cells. 
\subsection{ERBB-RAS-MAPK system and simultaneous signal measurement by the fluorescence microscope}
\begin{figure}[t]
  \centering 
  \includegraphics[width = 8cm]{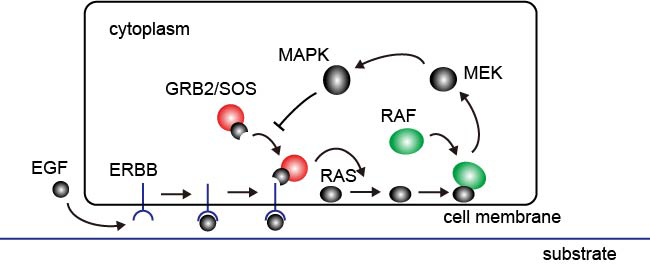}
  \caption{ Signal transduction pathway of the ERBB-RAS-MAPK system. In cells stimulated with EGF, successive translocations of SOS and RAF from the cytoplasm to the cell membrane occur, which recognizes ERBB and RAS activation, respectively. The membrane associations of SOS and RAF and RAF-induced MAPK activation create a negative feedback loop, which is disrupted by the R1131K mutation of SOS. See the text for details. 
  \label{fig4}}
\end{figure}

The ERBB-RAS-MAPK system is an intracellular signal transduction network responsible for cell fate decisions \cite{Lemmon2010}. ERBB is a cell surface receptor protein activated by small proteins, including epidermal growth factor (EGF), applied to the cell culture medium. 
%The ERBB activation is recognized by the GRB2/SOS protein complex in the cytoplasm meaning association 
% of the complex to 
%with the ERBB at the cytoplasmic side of the cell membrane (figure \ref{fig4}). 
{ The ERBB activation is recognized by the GRB2/SOS protein complex in the cytoplasm and is associated with the ERBB at the cytoplasmic side of the cell membrane (Figure \ref{fig4}).}
Subsequently, SOS activates RAS protein on the membrane to induce an association of RAF protein in the cytoplasm with the active RAS. The recruitment of SOS and RAF to the cell surface can be detected under a fluorescence microscope, reflecting the activation of ERBB and RAS, respectively \cite{Nakamura2017}. SOS and RAF can be observed simultaneously in the same single cells using two different colors of fluorescent tags \cite{Yoshizawa2021}.

\begin{figure}
\centering
   \includegraphics[width = 8cm]{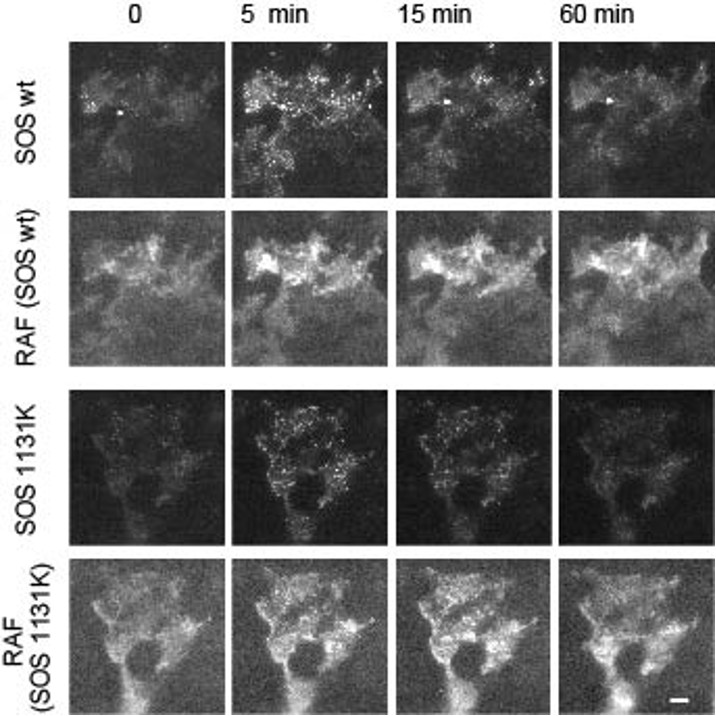}
  \caption{ Recruitments of SOS and RAF to the cell membrane in cells stimulated with EGF. Fluorescence signals from the basal surface of the cells were selectively observed using a total internal reflection fluorescence microscope. At time 0, cells were stimulated with EGF. The upper (SOS) and lower (RAF) images were acquired at the same field of view using a dual color microscopy. Scale bar: 10 $\mu$m.
  \label{fig5}}
\end{figure}

%\begin{figure}[h]
%  \centering
%
%\end{figure}

\begin{figure}[t]
  \centering
  \includegraphics[width = 10cm]{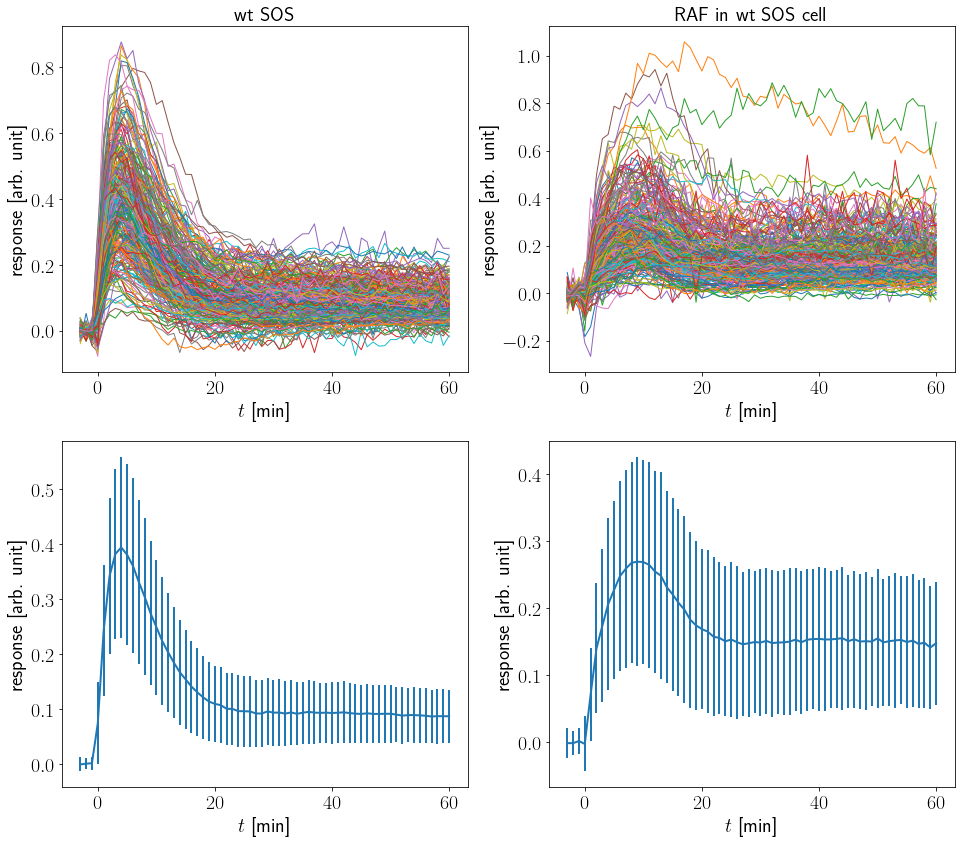}
  \caption{ Top panels: 248 samples of raw signals measured from cells with wt SOS. 
  Bottom panels: Their averages together with one standard deviation. 
  Left ((a), (c)) and right ((b), (d)) panels correspond to SOS and RAF, respectively. 
  \label{fig6}}
  \end{figure}
  \begin{figure}[t]
\centering
\vskip\baselineskip
   \includegraphics[width = 10cm]{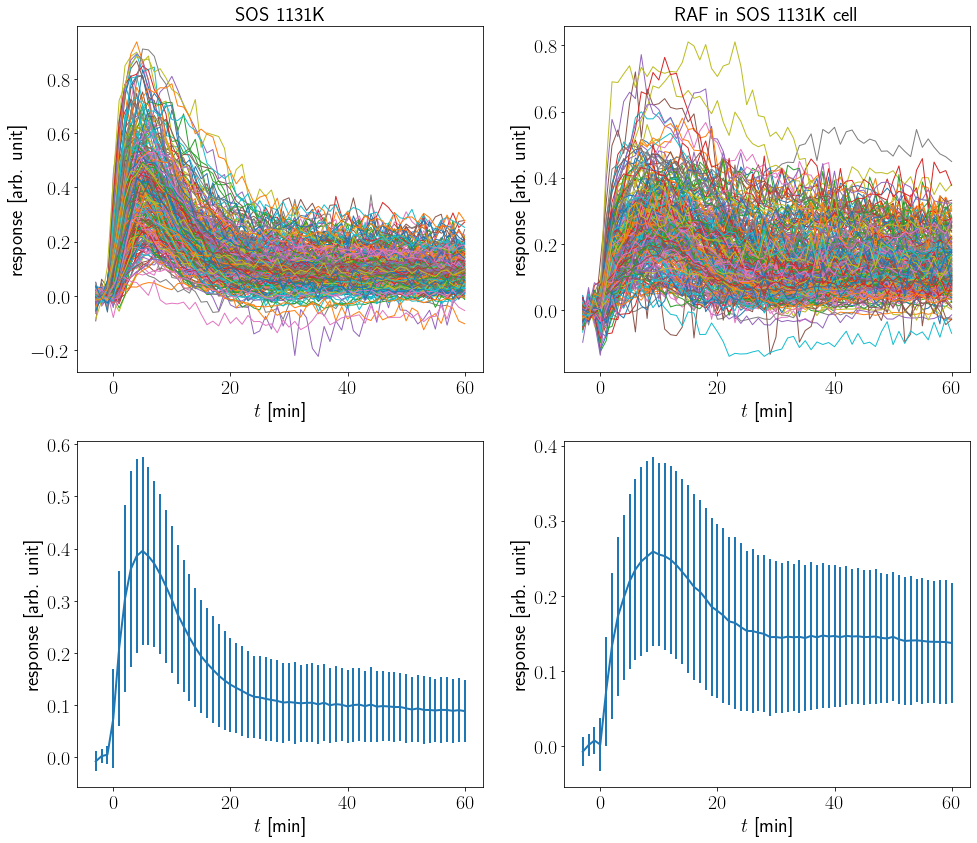}
  \caption{ Top panels: 282 samples of raw signals measured from cells with mutant (R1131K) SOS.  
  Bottom panels: Their averages together with one standard deviation. 
  Left ((a), (c)) and right ((b), (d)) panels correspond to SOS and RAF, respectively. 
  \label{fig7}}
\end{figure}

\subsection{Experiment and measured signals}	
	We stimulated ERBB in HeLa cells expressing the wild-type or mutant (i.e., R1131K) SOS with EGF under a microscope {at time $t=0$} and measured the changes in the fluorescence signals from SOS and RAF on the cell membrane (Figure \ref{fig5}). R1131K is a mutant of SOS, in which the arginine (R) at position 1131 in the amino acid sequence is replaced by lysine (K). This mutation is observed in Noonan syndrome, a human genetic disease \cite{Lepri2011}. The hyperactivation of RAS has been reported under this mutation, but its molecular mechanism is not entirely known. One possibility is the defect of a negative feedback regulation caused by serine phosphorylation around R1131, which is in the GRB2 association site of SOS \cite{Nakamura2017,Corbalan-Garcia1996}. 

	Figures \ref{fig6} and \ref{fig7} show 248 and 282 samples of the signals measured from 
	cells with wild-type (wt) or mutant SOS, respectively. The unit time is 1 minute, and the signals represent
	the increase of the fluorescence intensity from the average levels of the first three points corresponding to 
	$t= -3, -2$, and $-1$ [min]. 
As shown in these figures, transient increases of the fluorescence signals of SOS and RAF are observed after EGF application. In addition, the RAF activation dynamics are delayed and sustained after the initial peak. The difference in the activation dynamics between cells with wt and mutant SOS is barely noticed in the average and single-cell trajectories.
	
\begin{figure}[t]
  \centering
  \includegraphics[width = 8.7cm]{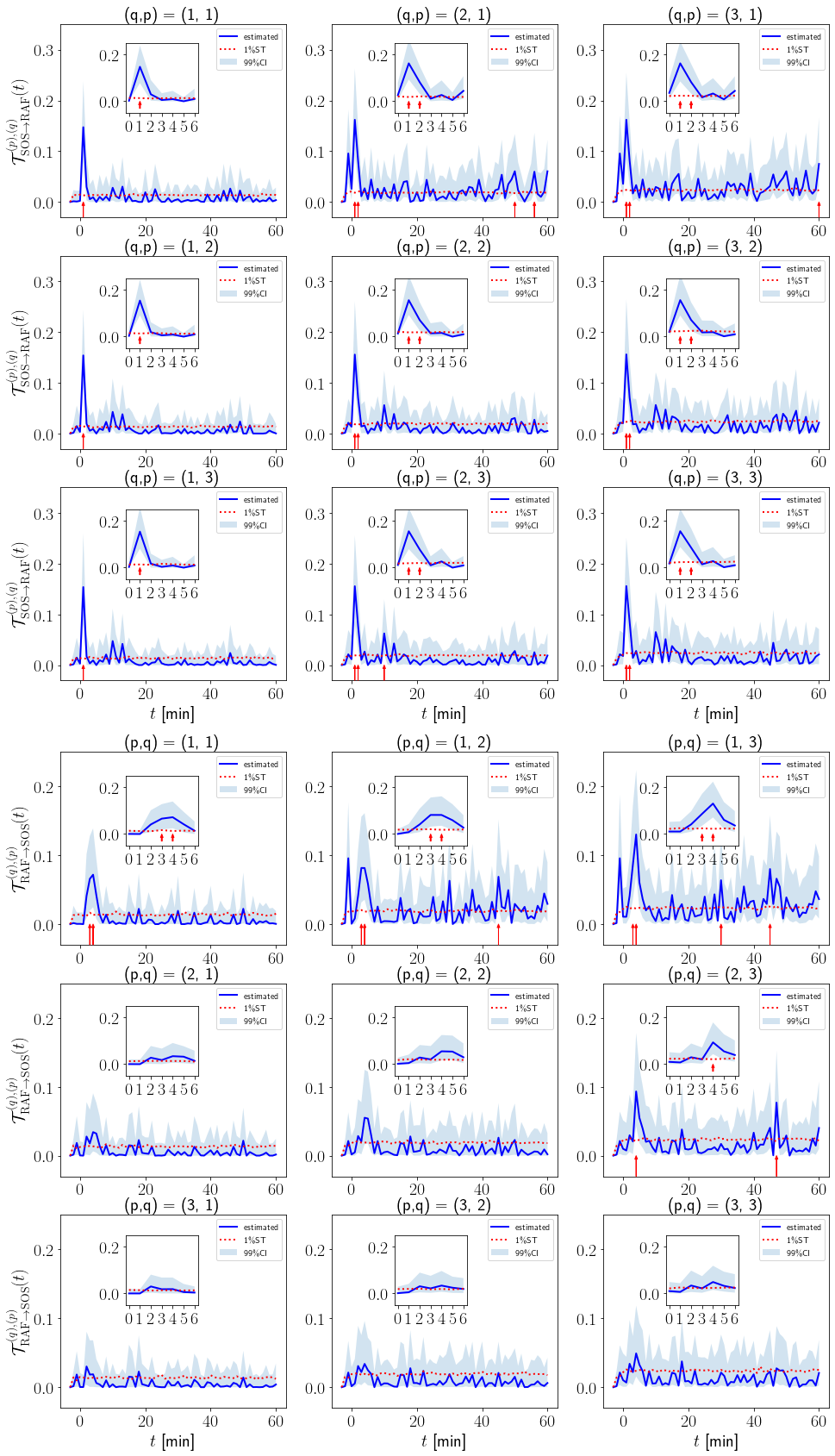}
  \caption{ Assessed TE (blue full lines) for cells with wt SOS from samples shown in Figure \ref{fig6}. 
  (a): From SOS to RAF. (b): From RAF to SOS. 
  %%%%%%
  Light blue areas and red dotted lines represent 99\%CIs and 1\%STs, respectively. 
  %%%%%%%
  The red arrows indicate statistically significant estimates, 
  which are screened %by setting the statistical confidence and the significance levels to 99\% and 1\%, respectively.  
  by the statistical confidence and the significance levels. 
  In response to the initial peak of TE from SOS to RAF, the significant TE in the reverse direction was found after 2--3 min (insets). 
  Statistically significant TE is also observed even in the later stage. 
  \label{fig8}}
\end{figure}

\begin{figure}[t]
  \centering
  \includegraphics[width = 8.7cm]{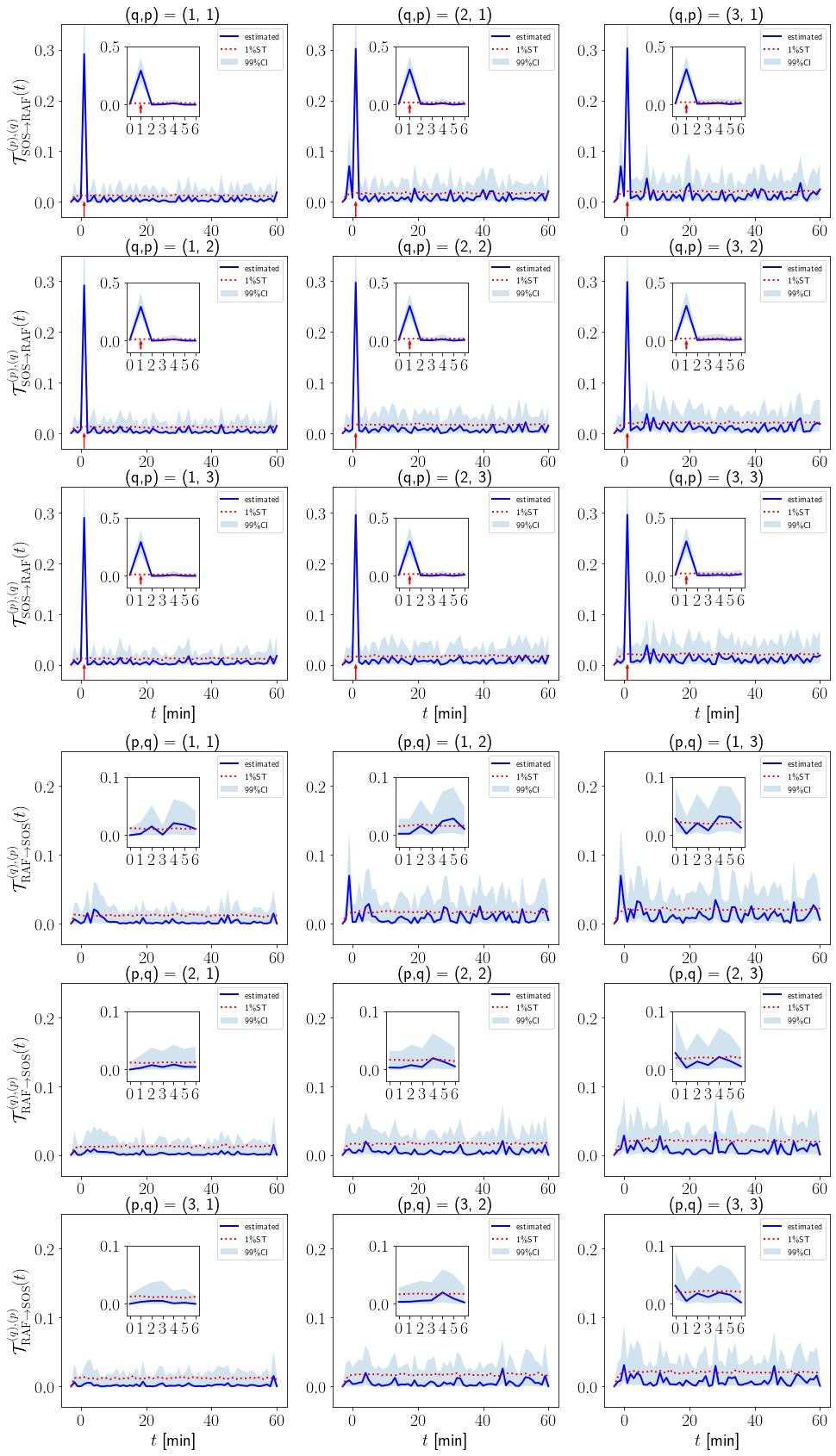}
  \caption{ Assessed TE (blue full lines) for cells with mutant SOS from samples shown in Figure \ref{fig7}.
   (a): From SOS to RAF. (b): From RAF to SOS. 
   %%%%%%
  Light blue areas and red dotted lines represent 99\%CIs and 1\%STs, respectively. 
  %%%%%%%
   The red arrows indicate statistically significant estimates, 
  which are screened %by setting the statistical confidence and the significance levels to 99\% and 1\%, respectively.  
  by the statistical confidence and the significance levels.  
   No statistically significant TE was found in the direction from RAF to SOS, even in the early stages (insets).
  \label{fig9}}
\end{figure}

\subsection{Detecting the difference between cells with wild-type and mutant biomolecule by TE}		
Figures \ref{fig8} and \ref{fig9} show the assessment of TE values for the samples presented in Figures \ref{fig6} and \ref{fig7} by changing the lags $p$ and $q$ systematically. 
{\color{black} A code and raw data for reproducing these figures are available from a web site\cite{code_url}.}
The sample sizes of 248 and 282 for cells with wt and mutant SOS, respectively, are considered not to be too small 
compared with the degrees of freedom $(p+q+2)(p+q+1)/2$ of the covariance matrices to be estimated within the range of $1\le p\le 3$ and $1\le q\le 3$.
For all cases, the number of bootstrapping repetitions $B$ for assessing CI and ST was set to 1000. 

Significantly large values of TE from SOS to RAF 
are observed at the early stage of cell signaling in all $(p,q)$ combinations for both wt and mutant SOS, which is expected from the signal transduction cascade (Figure \ref{fig4}). 
However, from RAF to SOS, significant TE values are obtained only in cells with wt SOS (Figure \ref{fig8}). 
The initial peaks of TE from RAF to SOS delay
by approximately 2--3 minutes
from those from SOS to RAF (insets), suggesting reversed information flow caused by the negative feedback loop 
from RAF to SOS via MAPK and/or other proteins downstream of RAF. 
In addition, the initial peak of TE from RAF to SOS increases as the lag $q$ of the RAF is set larger, 
while that of the reverse direction does not exhibit such a tendency.  
The result of the theoretical model in Section 2 implies that this may be because 
there is a longer cascade of reactions in the signal transduction pathway from RAF to SOS 
than in that from SOS to RAF. 
{The first peak values of TE from RAF to SOS decrease as $p$ increases,
while those from SOS to RAF hardly depend on $q$.  
This may be due to the difference of noise levels in the measurement between the SOS and RAF. 
%In this study, the noise levels are affected by the nature of the fluorescence tag bound to the proteins. 
%We used tetramethylrhodamine, which is a chemical probe that is brighter and more stable than fluorescent proteins such as GFP, and GFP for SOS and RAF, respectively. 
The noise levels in our measurement are affected by the nature
of the fluorescence tag bound to the proteins. We used tetramethylrhodamine and GFP
for SOS and RAF, respectively. The former is a chemical probe that is brighter and
more stable than fluorescent proteins like GFP.
}

Another striking feature is that statistically significant TE values are observed in cells 
with wt SOS at the later stage, while no such behavior is found in cells with mutant SOS. 
As RAF activation is sustained in cells with SOS of both types (Figures \ref{fig6} and \ref{fig7}), 
this difference may be a collateral evidence of the defect of a negative feedback regulation from 
RAF to SOS in cells with mutant SOS.

%%%%
{\color{black}
Under the Gaussian approximation, TE can be computed from covariances of two time series with 
given time lags. This may invoke a naive question whether similar results can also be obtained by 
more conventional covariance based analyses. 
For answering such a question, we plot covariance and Pearson's correlation coefficient, 
which is defined by normalizing the covariance with product of standard deviations, 
between SOS and RAF changing time lags in Figures \ref{fig10} and \ref{fig11} for cells with wild-type and mutant SOS, respectively. 
The plots show that the two time series with the time lags are correlated with the statistically significant level 
in the almost entire range of observation time regardless of whether the SOS type is wild or mutant. 
Although the profiles of peaks and the strength of correlations differ slightly, 
it is difficult to find distinct qualitative difference between the two cases from the plots. 
This implies that TE is a more suitable measure for characterizing the information flow
conveyed between the two time series with a high time resolution. 
	
\begin{figure}[t]
  \centering
  \includegraphics[width = 9cm]{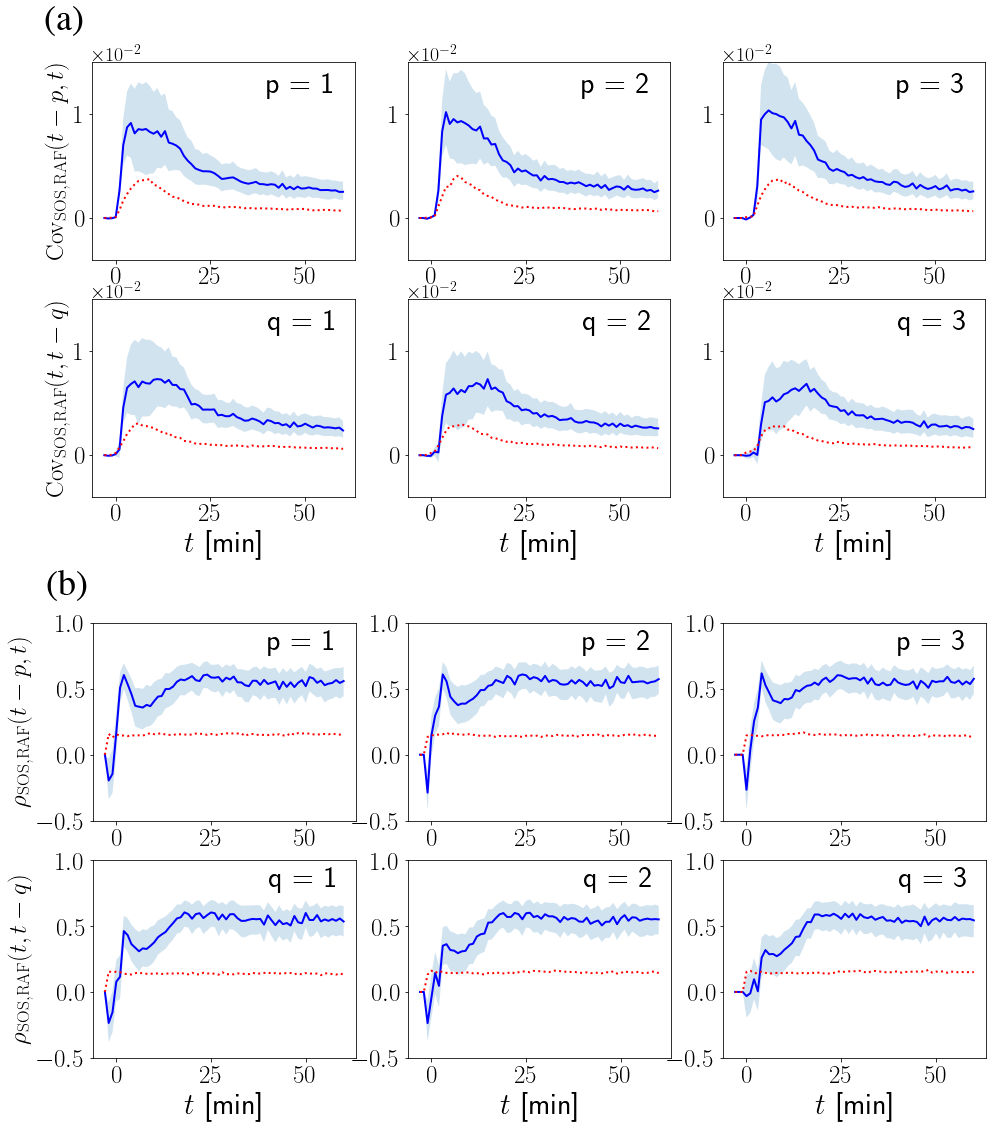}
  \caption{ (a): Covariance and (b): Pearson's correlation coefficient between 
SOS and RAF activities with time lags for cells with wild-type SOS (blue full lines). 
99\% CIs (light blue areas) and CTs (red dotted lines), which represent one percentile point from the top for the null hypothesis, 
are computed by the bootstrapping method. 
  \label{fig10}}
\end{figure}

\begin{figure}[t]
  \centering
  \includegraphics[width = 9cm]{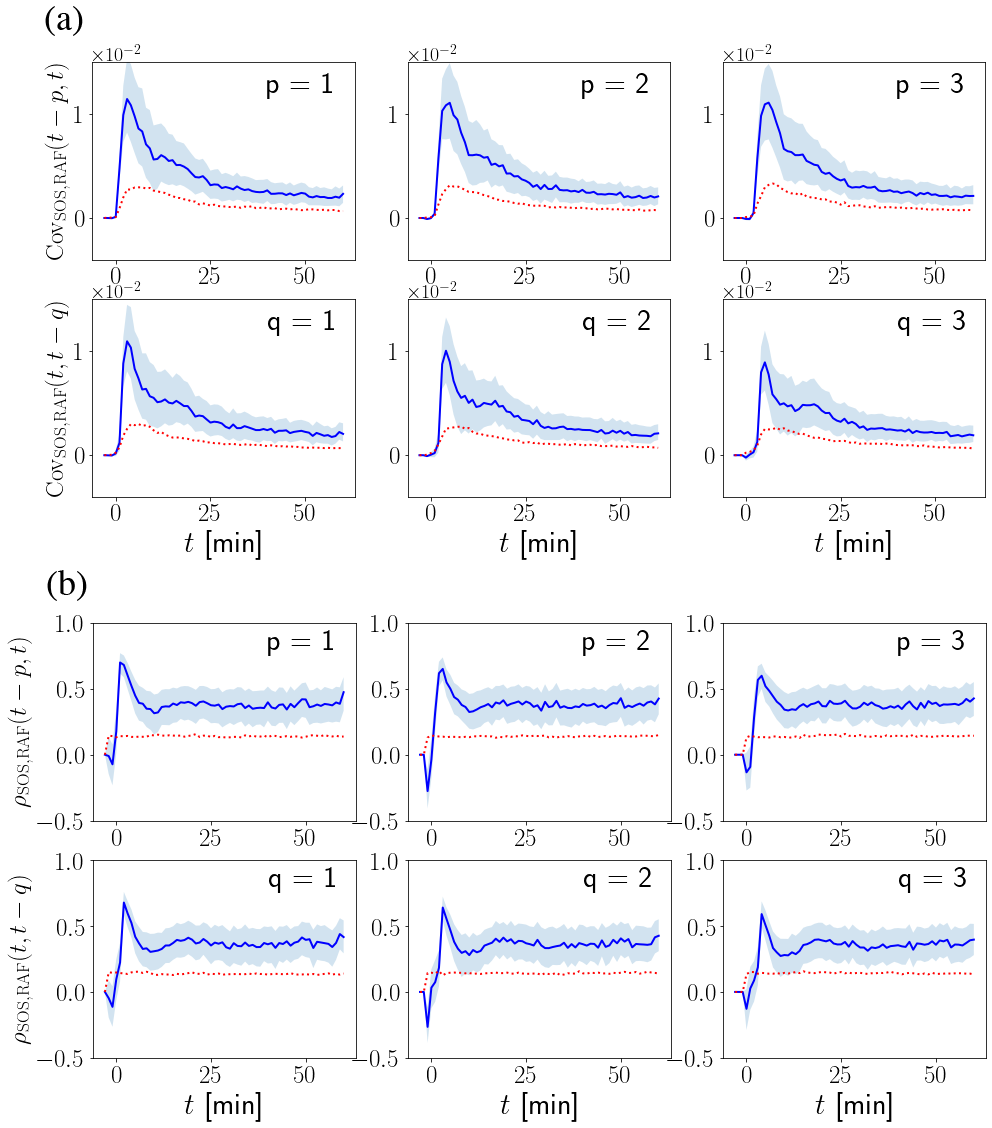}
  \caption{ (a): Covariance and (b): Pearson's correlation coefficient between 
SOS and RAF activities with time lags for cells with mutant SOS (blue full lines). 
99\% CIs (light blue areas) and CTs (red dotted lines), which represent one percentile point from the top for the null hypothesis, 
are computed by the bootstrapping method. 
  \label{fig11}}
\end{figure}
}

\section{ Summary and discussion}
In summary, we examined the possibilities and limitations of assessing the transfer entropy (TE) from the measured data 
of biochemical reactions. We employed the Gaussian approximation,  
which enables us to efficiently assess TE based on covariance matrices estimated from samples of objective time series. 
In general, it is necessary to evaluate the sampling distributions to guarantee the accuracy of the estimated results. 
However, an analytical evaluation of the sampling distributions of TE is difficult for non-stationary time series. 
We resolved this difficulty by computationally assessing the sampling distributions
using bootstrapping techniques from computational statistics. 
The computational methods were tested by the application to a theoretically 
tractable model of a stochastic process, 
which led to the development of a method for screening only statistically significant estimates under given 
levels of statistical confidence and significance. 
In addition, this method was applied to assess the dynamics of the information flow in a real biological reaction network inside living cells. 
Although the raw signals measured from cells with the wild-type and a mutant molecule are hardly distinguished, 
the method successfully detected the difference between them in the time course of TE. 
This implies that the developed method may serve as a useful tool in studying intracellular reaction networks in which large scale simultaneous measurement of activities of biomolecules has been made possible owing to the recent advancement of fluorescence microscope technologies.

%%%%%
{\color{black}
It should be noted that the finiteness of TE is not a sufficient sign of the causal relationship between 
two time series \cite{Baldovin2020}; non-zero TE can be a spurious causality detector when observations are performed 
incompletely due to the presence of unobserved states \cite{Smirnov2013}. 
Nevertheless, the TE based analysis shown in the current paper 
would still be useful at least for the following two purposes. 
One is to quantify the efficiency of information transmission for {\em known pathways} for which 
unobserved states are absent or their influence is negligible.  
As mentioned in Introduction, many pathways of biochemical reactions have been identified with a high accuracy for these decades. 
However, little is known about when and how large information is transmitted through the pathways. 
Our methodology could fill the ``missing piece'' with a high time resolution
although hot debates are still being continued on the appropriateness of 
TE as a causality quantifier \cite{Smirnov2014,Stokes2017,Dhamala2018,Barnett2018,Smirnov2020a,Smirnov2020b}. 
The other is to offer clues for finding {\em unknown pathways}.  
Although the finiteness of TE is not a sufficient condition of 
the causal relationship, it still serves as a necessary condition. 
Therefore, the assessment of TE would provide useful guidelines for screening possible candidates 
of relevant pathways.  
In addition, even if pathways are not identified completely, comparison of TE between healthy and disordered systems might lead to 
more efficient diagnosis and treatment of various disorders.  
}

%%%%%
Although we restricted the application domain to biochemical reactions in this study,  
the proposed methodology can be utilized to analyze information flow in general systems of a wider class. 
%%%%%
{\color{black}
For instance, it may be useful in examining time varying effective interactions in multi-agent/non-linear dynamical systems 
from simulation data \cite{Vicsek1995,Andrzejak2006,Cavagna2014} and those in nervous/active matter systems 
from video imaging data \cite{Ota2021,Iwasawa2021}, 
as phenomena observed in such systems are fairly reproducible and collecting many data on them is relatively easy. 
%%%%%
Nevertheless, 
the necessity of collecting many samples of simultaneously measured data (i.e., at least hundreds of samples) 
for accurately estimating covariances may be a bottleneck in applying it to non-stationary time series in %the 
many other domains. }
Meanwhile, in general, appropriate prior knowledge about objective systems can improve the estimation
accuracy significantly. 
The reduction of the necessary sample size incorporating Bayesian inference and/or other machine learning techniques
is an important research direction for further research. 

\section*{Acknowledgments}
We acknowledge the technical assistance provided by Mutsumi Nakanishi. 
This work was partially supported by the MEXT KAKENHI Grant No. 19H05647 (YS),  
JSPS KAKENHI Grant Nos. 17H00764 (TO, YK), 18K11463, 19H0182 (TO), 20H00620 (YK), and JST.
CREST Grant No. JPMJCR1912 (YS, YK).

\appendix

{\color{black}
\section{Derivation of equation (\ref{GaussianEntropy})}
\label{GaussianEntropyDeriv}
For two random variables $\bm{X}$ and $\bm{Y}$ that follow joint distribution $Q(\bm{x}, \bm{y})$, 
where $\bm{X}$ and $\bm{Y}$ may be either scalar or any dimensional vector random variables, 
conditional entropy of $\bm{X}$ given $\bm{Y}$ is defined generally as
\begin{eqnarray}
H(\bm{X}|\bm{Y}) 
= -\int d\bm{y} Q(\bm{y}) d\bm{x}Q(\bm{x}|\bm{y}) \ln Q(\bm{x}|\bm{y}).
\label{conditional_entropy}
\end{eqnarray}
Let us suppose that $\bm{X}$ and $\bm{Y}$ follows a multivariate Gaussian, i.e., 
\begin{eqnarray}
&&Q(\bm{x}, \bm{y}) = \frac{1}{(2\pi)^{\frac{n+m}{2} }({\rm det}\Sigma(\bm{X}\oplus \bm{Y}))^{1/2}} \cr
&&  \phantom{Q(\bm{x}, \bm{y}) = }
\times 
\exp \left (-\frac{1}{2}(\bm{x}\oplus \bm{y} -\bm{\mu}_X \oplus \bm{\mu}_Y)^\top \Sigma(\bm{X}\oplus \bm{Y})^{-1} (\bm{x}\oplus \bm{y} -\bm{\mu}_X \oplus \bm{\mu}_Y) \right ), 
\label{GaussianQ}
\end{eqnarray}
where $n$ and $m$ are the dimensions of $\bm{X}$ and $\bm{Y}$, and 
$\bm{\mu}_X$ and $\bm{\mu}_Y$ are means of $\bm{X}$ and $\bm{Y}$. Covariance matrix $\Sigma(\bm{X} \oplus \bm{Y})$ is expressed as
\begin{eqnarray}
\Sigma(\bm{X} \oplus \bm{Y}) = \left (
\begin{array}{cc}
\Sigma(\bm{X}) & \Sigma(\bm{X}, \bm{Y})\cr
\Sigma(\bm{Y}, \bm{X}) & \Sigma(\bm{Y})
\end{array}
\right )
\label{covariance_decompose}
\end{eqnarray}
using the notation defined in the main text. 

Computing the matrix inversion of (\ref{covariance_decompose}) yields an expression 
\begin{eqnarray}
\Sigma(\bm{X} \!\!\oplus \!\!\bm{Y})^{-1} = \!\!
\left (
 \!\!
\begin{array}{cc}
\Sigma(\bm{X}| \bm{Y})^{-1} & -\Sigma(\bm{X}| \bm{Y})^{-1}\Sigma(\bm{X}\!, \!\bm{Y}) \Sigma(\bm{Y})^{-1}  \cr
-\Sigma(\bm{Y})^{-1} \Sigma(\bm{Y}\!, \!\bm{X}) \Sigma(\bm{X}| \bm{Y})^{-1} & \Sigma(\bm{Y}|\bm{X})^{-1}
\end{array}
\!\! 
\right )
\end{eqnarray}
This means that conditional distribution $Q(\bm{x}|\bm{y})$ is expressed as
\begin{eqnarray}
&&Q(\bm{x}|\bm{y}) = \frac{1}{(2\pi)^{n/2} {\rm det} (\Sigma(\bm{X}|\bm{Y}))^{1/2}} \cr
&& \phantom{Q(\bm{x}|\bm{y}) =} \times \exp\left (-\frac{1}{2} (\bm{x} - \bm{\mu}_{X|Y})^\top \Sigma(\bm{X}|\bm{Y})^{-1} (\bm{x} - \bm{\mu}_{X|Y}) \right ), 
\label{conditional_dist}
\end{eqnarray}
where $\bm{\mu}_{X|Y} = \bm{\mu}_X +\Sigma(\bm{X}, \bm{Y}) \Sigma(\bm{Y})^{-1} (\bm{y} -\bm{\mu}_Y)$. 
Inserting this expression into 
(\ref{conditional_entropy}) offers
\begin{eqnarray}
H(\bm{X}|\bm{Y}) = \frac{1}{2} \ln {\rm det}\left (\Sigma(\bm{X}|\bm{Y}) \right ) +\frac{n}{2}\ln (2 \pi e). 
\label{GaussianConditionalEntropy}
\end{eqnarray}
Finally, substituting  $X_t$ and $\bm{X}_{t-1}^{(p)}$ with $\bm{X}$ and $\bm{Y}$, respectively, in (\ref{GaussianConditionalEntropy}) provides (\ref{GaussianEntropy}). 
}

%\bibliography{information_flow} 

%aipnum4-2.bst 2019-01-14 (MD) hand-edited version of apsrev4-1.bst
%Control: key (0)
%Control: author (8) initials jnrlst
%Control: editor formatted (1) identically to author
%Control: production of article title (0) allowed
%Control: page (1) range
%Control: year (1) truncated
%Control: production of eprint (0) enabled
%

\end{document}